Consciousness Viewed in the Framework of Brain Phase Space Dynamics, Criticality, and the Renormalization Group.

Gerhard Werner
Department of Biomedical Engineering
University of Texas at Austin

Abstract

   To set the stage for viewing Consciousness in terms of brain phase space dynamics and criticality, I will first review currently prominent theoretical conceptualizations and, where appropriate, identify ill-advised and flawed notions in Theoretical Neuroscience that may impede viewing Consciousness as a phenomenon in Physics. I will furthermore introduce relevant facts that tend not to receive adequate attention in much of the current Consciousness discourse. As a new approach to conceptualizing Consciousness, I propose considering it as a collective achievement of the brain's complex neural dynamics that is amenable to study in the framework of state space dynamics and criticality. In Physics, concepts of phase space transitions and the Renormalization Group are powerful tools for interpreting phenomena involving many scales of length and time in complex systems. The significance of these concepts lies in their accounting for the emergence of different levels of new collective behaviors in complex systems, each level with its distinct ontology, organization and laws, as a new pattern of reality. The presumption of this proposal is that the subjectivity of Consciousness is the epistemic interpretation of a level of reality that originates in phase transitions of the brain-body-environment system.

1. Introduction.

   The metaphysical tradition, prevailing in the West since the 17th Century, views reality as objective in the sense of being accessible and, in principle, knowable to all observers. The study of mental phenomena such as beliefs, desires, thoughts, hopes, fears and conscious states of subjective experience, generally, conflicts with this position: it excludes the region of reality to which these subjective phenomena belong. Yet, as we know well from personal experience, these phenomena do exist. To accommodate this fact, Searl [122] proposed that the phenomena of subjectivity are realized in the brain as an Ontology to which we have epistemic access in the form of the features of our subjectivity. Expanding on this view, the subjectivity of the Mental is not merely an epistemological fact; rather the epistemic access points towards a (physical) ontology whose intrinsic properties ( ontic states ) are epistemically accessible as the features of subjectivity (for ontic and epistemic states, see Section 3.3). Thus, the central goal of this essay is to introduce a point of view that the Ontology of the Subjective is constituted within the framework, and under the auspices, of the laws of Physics and Biology. Emphatically, this is not to advocate a "naturalistic dualism" in the manner envisaged, for instance, by Chalmers [30] for closing the gap between consciousness and the ontology of natural sciences. Rather, it takes into account that biological systems generally [105] and brains in particular (for a recent review [31]) are complex dynamical systems poised at criticality. In this case, levels of observation (and, thus, reality) are related by phase transitions which are the subject of critical phenomena in complex systems of non-equilibrium statistical mechanics [61, 140]. They are known to lead to the collective emergence of multiple levels of organization, each with its own ontology, and epistemic descriptors as the knowledge that is empirically available about a physical system's intrinsic properties.

   A specific aspect of this point of view is the idea of the Renormalization group [67] which considers Reality composed of a hierarchy of levels, related to one another by phase transition, with each level representing a new ontology as a (qualitatively) new pattern of Reality [112]. I suggest that the Ontology of the Subjective is one of these levels of Reality, arising in the Renormalization Group transformation of the brain's phase space dynamics; and that the epistemic access to it constitutes our Subjectivity. For earlier publications discussing aspects of this line of thought , see [165, 166]. The contention is that the structural features of the Ontology of Subjectivity correspond to the structure of consciousness, of which some of the phenomenological characteristics are its unity, the varying degree to which it can be associated with other mental faculties, its association with intentionality, and self-attribution and reportability of mental states, to list but a few. To sidestep the constraints of Phenomenology, the proponents of Neuroscience studies of consciousness proposed several lists of observationally testable indicators for consciousness in humans [52, 121, 126, 129]. Contravening the charge of anthropocentrism of consciousness studies, G.M. Edelman et al [51] and D.B. Edelman and Seth [50] discussed a programmatic framework for determining necessary conditions that would ascertain features of consciousness in non-mammalian species: avian species passed the test, but the case for Cephalopodes remains undecided. Seth [124] asks the important question: " Does Consciousness have a function ? and if so, what might it be "? Interest in this question is, in part, due to seeking guidance for designing artificial consciousness. In view of consciousness not being a unitary phenomenon, it is useful to distinguish a primary consciousness for ongoing perceptual-motor adaptability in the present from the higher order consciousness of self-reflection and interpretative functionality. In any case, however, the capacity for globally mobilizing and integrating separate and independent functions seems to be a distinctive property [11].

Notice how decisively the proposed outlook differs from the approaches to the 'Mind Body' relation that seeks to ground subjective phenomena (in part also those of Folk Psychology) reductively in a biological-physical ontology. The inadequacy of this attempt is, in part, evident from the strictly internalist stance of the "New Wave Reductionism " [25,156] of rising popularity. Internalism is also a liability in considering neural and physical processes or states as correlated [109]. Still more damaging, however, is the confounding of correlation with causation. The notion of "Neural Correlates of Consciousness" proposed by Koch [86, 148]) as the " minimal neuronal mechanisms jointly sufficient for any one specific conscious percept" is at best of limited value, at worst misleading since correlation must not be taken as evidence of causality except in the context of an appropriate theoretical framework [5,57,58]. But this is missing. In distinction, the *Explanatory Correlates* of Seth [123] intended to account for essential features of consciousness, meet this requirement by being anchored in existing theories. By virtue of this, they can also set the stage for simulation models and possibilities for designing conscious artifacts [35]. The discussion of these issues will be taken in the following sections.

3. The background

3.1 On current neurobiological theories of consciousness

In general terms, many current theories of brain function attribute a significant role to neuronal synchrony at a meso- or macroscopic level. In the wake of Varela's et al [158] publication on the Brain Web, the coordination of activity between neurons and neuron assemblies by phase synchrony of neuron firing patterns became an intensely studied research topic, motivated in part by seeking to ascertain its potential relevance for performance in psychophysical tests and , possibly, also for consciousness: citing merely a few recent reviews: Varela and Thompson [157], Melloni et al [98], Uhlhaas et al [155]. A collection of papers in "Dynamic Coordination in the Brain: from Neurons to Mind" [159] also discusses these and various related aspects. Consequences of synchronous activity of neurons in the form of neuronal discharge patterns, membrane potentials or as field potentials may be variously described in terms of coupling or coordinating activity, implying some form of information sharing. However, synchrony does not necessarily capture the totality of informational sharing that may obtain, nor do for various reasons, cross correlations or mutual information, as Klinkner et al [84] proved. A new measure of 'Informational Coherence' for estimating the neural information sharing and coordinated activity, designed by these authors, is based on mutual information of dynamical states (constructed as causal state models), rather than merely observables. The analysis by Klinker et al warrants some reluctance to accepting phase synchrony as meaningful measure, and Tononi and Laureys [147] raise additional arguments for skepticism regarding the intuitively appealing, though conceptually missing link between neuronal synchrony (or its informational equivalent) on the one hand, and the "binding" of elements of neural activity to unified percepts.

Among neurobiological theories of consciousness, (see, for instance [87][ , I turn in the following to the framework of the Neuronal Group Selection theory [ 52,53] because of its detailed formulation, and the multiple ramifications it engenders. It features prominently the the dynamic interaction among widely distributed groups of neurons via reentrant and reciprocal mapping [52], Among those interactions, the thalamo-cortical system is thought to occupy a privileged role as the 'dynamic core'. Performance in computational simulations are interpretable in terms of generally recognized properties of consciousness [129]. The notion of the reentrant dynamic is also readily compatible with the imaginative synthesis of evidence from Psychology, which Baars [14] introduced as the ' Global workspace theory' of conscious experience. Baars developed its principles subsequently to a 'cognitive theory of consciousness' [11,12,13] according to which multiple processors dynamically constitute context dependent coalitions for gaining access to a limited capacity global workspace. This principle has become a central component in the Neuronal Group framework as well as related constellation of ideas. Remarkably, a robot designed on the information flow principles of this theory displayed anticipation and planning based on internal simulation of interactions with the environment, as well as action selection mediated by an affect-like valuation function [132]. I will return to this issue in Section 3.3. In a separate and extended series of studies, Dehaene, Changeux and colleagues [44,45] implemented global workspace architectures as "neuronal global workspace" for comparison with performance in psychophysical tests, and to dissect conscious, preconscious and subliminal forms of processing [43,46]. For a more detailed account of this remarkable convergence of foundational ideas of varied origin and nature, I refer to my previous review [168]. Finally, Wallace [161,162] contributed two elegant mathematical theories of the global workspace idea to which I will return in Sections 4 and 5.

The dynamic core theory engendered seminal ideas designed to probe more deeply some aspects of its functional repertoire: one of them is the definition of a complexity measure for the brain [153, 154]. The approach rests on the intuition that cognition would require integration of multiple disparate sources (subsystems) in the brain [52]. Complexity, in the present context means interdependence among subsystems. It can be estimated as the statistical measure of Mutual Information shared by subsystems: accordingly, their independence is reflected in a low, and their integration in a high value of this measure. Computational models featuring reciprocal and parallel connections among functionally segregated groups of neurons exhibit spontaneously high values of the Mutual Information shared by the constituent neuron groups, as do cortical connection matrices based on neuroanatomical data from macaque visual cortex, implemented as dynamical systems [138, 139]: these cortical connection patters do indeed generate functional connectivity of high complexity, based on highly connected and coherent functional clusters. Accordingly, the proposed complexity measure does capture the degree of subsystem integration which can be attributed to correlated patterns of neuronal activity among different groups of neurons. Moreover, the value of the complexity measure changes with stimulus induced alterations of functional connectivity, whereby the model's intrinsic complexity comes to match adaptively the statistical structure of the external input. Subsystem integration, reflected in Mutual Information, is in this setting thought of as a form of Information processing [152]. However, it must be noted that the proposed complexity measure is not unique: it is just one of a family of information-theoretic metrics based on the intuition of segregation-integration balance [18]. I will resume this point when discussing an extended framework of theories and measures of consciousness [128].

Proceeding from the foregoing basis, Tononi and associates [16, 149, 150, 151] formalized and refined in successive steps the "Information Integration Theory of Consciousness". At the fundamental level, consciousness is in this theory viewed as integrated information: 'Integrated information" is understood as information generated by a system selecting *one* from among all possible states it can assume. The particular state selected is a function of two factors: the system's range of *a priori* available states, and those states that the system *a posteriori* (i.e. after receiving some input stimulus) intrinsically, on the basis of its intrinsic architecture and dynamics, identifies as being causally related to elements of its a priori repertoire. The difference between *a priori* and *a posteriori* selection is the effective information that matters: Integrated Information is then the information generated (i.e. reduction of uncertainty facing a stimulus) attributable to causal interaction among system elements, in excess of the information generated by the system's parts. The theory postulates that this quantity is equivalent to the level of consciousness. Given these assumptions, it is then possible to quantify Integrated Information in models as function of system architecture and dynamics: although for computational reasons limited to small system size. Tononi et al show in support of this theory that neurobiologically plausible system architectures are associated with high levels of Integrated Information. Balduzzi and Tononi [16] then go on to develop mathematical procedures for characterizing informational relations for dynamical systems of discrete elements which evolve in state space by Markovian transitions. Information Integration is in this model measured as the information generated by the system's transition to one particular out of the possible states. The reasoning is thus analogous to the earlier described (static) measure, but based on evolving shapes in the system's state space, rather than on static values. Balduzzi and Tononi [16] then argue that these shapes meet criteria for characterizing qualities of conscious experiences. An initial attempt to model essential aspects of the theory succeeded in generating meaningful behavior in a virtual robot [66]. Nathan and Barbos [107] apply a network algorithimic procedure to a computational model of the cortex for obtaining a measure of information integration. Induction of anesthesia reduces the brain's information integration capacity [93].

The essential point is the thought that a system generates more information than the sum of its parts, and that "integrated information" measures the extent to which the system's capacity allows this to occur, be it under static or dynamic regimes, associated with level versus quality of information (consciousness?), respectively. Barrett and Seth [19] challenge the approach chosen by Tononi and associates on two counts: first, because of the practical limitations of the proposed algorithms ;and, second, more fundamentally, because of its restriction to discrete Markov systems which limits the theory's generality. Barrett and Seth (l.c.) frame information integration in terms of process (rather than capacity; see above) and present an approach that would be suitable for simulating sensory-motor coordination in information rich environments.

In essential difference from Tononi's information-based approach, Seth and Edelman [127] and Seth [125] consider population activity patterns as causally effective rather than information bearing representations underlying computations. The essential elements of their implementation are graph theory and Granger causality (for details: [47, 69]) in the service of a principled causal connectivity analysis of network dynamics [130]: Granger causality, revealing recurrent structures of a network in which neurons are embedded, can be applied to represent interactions between variables as directed edges in graphs. In this form, the causal density of network dynamics is captured as the fraction of interactions among the nodes that are causally effective. Causal density is thus a process measure of network differentiation-integration. The Seth-Edelman model heeds the admonitions from the views on embedded Cognition (see for instance: [34]) by taking into account the continuous causal interactions among brain, body and environment. Behavioral learning via synaptic plasticity is viewed as shaping the selection of causal pathways in neural populations, in the framework of Edelman's Darwinian selectionist theory. These general notions are implemented, and put to test, in a brain-based device that allows the tracking of (simulated) neuronal activity during behavior in spatial navigation tasks in real environments ( Darwin X :[89], [90]) . The analysis centers on determining the multiple paths of functional interactions that lead in time to a selected neuron's activation with influence on behavior (the Reference Neuron, RN). The causal significance of every connection in the network of interacting paths leading to RN is estimated as Granger Causality , for each connection based on the activity time series of the pre- and postsynaptic neuron, respectively. Removing dead-end connections from the Granger network that do not participate in the causal chain leading to RN finally identifies its causal core.  Analyzing the model's patterns of activity revealed  several meaningful insights, of which I mention but two for the principled insight into  the network dynamics they afford:  variability of neural activity leading to a given event is the expression of the diversity of dynamic repertoires from among which selection occurs by pathway shifting in time; and  behavioral learning manifests itself at the population level as progressive refinement and eventual selection of (metaphorically speaking: by sculpting)  specific causal pathways (causal cores) from among the available large repertoire available for neuronal interactions.

In assessing merits and liabilities of the approaches  listed in the forgoing, Seth et al [128]  conclude that none of them fully capture the required multidimensional complexity of a neuro-behavioral system that could  account for objectively measurable features of consciousness; and that all of them have practical limitations. As landmarks for approaching satisfaction of these requirements, the authors continue to insist on adhering to the theoretical framework of the Neuronal Group Selection Theory and the concept of the dynamic core, emphasizing its participation in the transactional processes between organism (model) and the environmentally embedded body. The decisive issue lies in extending the previously considered notions of complexity. To this end, the notion of  a multidimensional  Relevant Complexity is introduced,  requiring at least  three dimensions of temporal, spatial and recursive complexity: the former two dimensions, each, covering a wide range of  temporal and spatial scales;  the later designating the integration-differentiation balance across different levels of system description which would ideally extend from the level of molecular synaptic dynamics to that of reentrant interactions among segregated brain regions. Implementing this  ambitious vision in a brain-based device would undoubtedly be a large step towards simulating behavior under conscious guidance.

Considering the idea of  Relevant Complexity as pivotal  requires paying tribute to the body housing the brain: following, in part,  the seminal insights of Damasio  [39]  the  global Workspace  theory acknowledges  the role of emotion and valuative judgments of Consciousness  [14]  taken already into account in Shanahan's [131] model, cited earlier.  However, it appears that the notion of Core Consciousness which Damasio formulates primarily on the basis of insights from Clinical Neurology warrants still much more detailed attention. The extensive studies of the Reticular Activating System of the neurophysiologist of the 50s and 60s , recently reviewed by Steriade [143]  and  placed in the context of sleep and arousal ([97],  and the elaborate mechanisms of registering the body's condition  [37]  assume renewed significance for  the somato-sensing functions  and structures, necessary for consciousness [110]. Evidently, an  extended infrastructure  subserves  interoceptive integration, orchestrating the  cortico-thalamic dynamics with the interests and capabilities of the body, and its homeostatic regulation, in the service of consciousness [38].

The next section will briefly sketch some snapshots of the prevailing conceptual landscape of  Cognitive Neuroscience, as part of the intellectual climate with potential influence on the orientation in  neurobiology-of-consciousness studies.

3.2: Consciousness "in the wild"

Paraphrasing  the title of Hutchins's [75]  well known book is intended to announce the target of this section which is to review, however briefly and not necessarily in any order of priority, what it takes to be conscious under conditions of ordinary, daily life. You need not be a fan of the criticism of neuroscience discourse leveled by Bennett and Hacker [22] , nor an ardent Wittgensteinian,  to acknowledge that the often encountered  locution "brains think" or "brains are conscious" is drastically wrong and misleading, even  if only used as  a shortcut or metaphorically. What IS to be considered conscious are persons, embedded and immersed in a material and social world in reciprocal interaction by which they are in large measure constituted (for a recent lucid analysis, see [120]). This does not of course preclude animals from having some forms of consciousness, as mentioned earlier, but for present purposes it is persons I am having in mind, whose capability for intelligent action includes the purposeful utilization of resources the environment offers in the form of sensory-motor interaction on which numerous authors agree [34, 100, 118, 170]. The extent to which the environment itself is in fact an 'extended mind" is at present controversial [32, 33, 99,163]

To some extent reminiscent of the submarine navigator in Maturana and Varela's "Tree of knowledge" [96] Metzinger [101, 102] developed in large detail a Self-Model Theory of Subjectivity (SMT). Comparing the function of the human brain with that of a flight simulator, it receives continuous input from sensory organs for constructing an internal representation of the external reality. Experientially, one is not aware that this internal representation is but a model of the external reality. Conscious experience and first- person perspective of an individual Self of Consciousness is said to consists in activating this world model, but leaves conditions for activation unexplained. In this world of virtual reality and embodied simulation, the motor system constructs goals, actions and intending selves [103], and enters into social traffic [63, 64],, the latter thought to involve specifically a class of neurons (appropriately called Mirro Neurons) in the forebrain for bi-directional interaction with the environment [116]. More generally, Gallese and Lakoff [65] suggest that the sensory-motor system has the right kind of structure for characterizing some abstract concepts: their idea of "neural exploitation" refers to putting the internalized model of sensory-motor brain mechanism in the service of new roles in Cognition. One should expect support of SMT from reports of out-of-body experience. The principle of these studies is to experimentally induce multisensory conflicts which would disrupt the pre-reflective bodily foundations of the experiential self-model [27]. Numerous observations do indeed attest that such conditions distort spatial unity, create the experience of a virtual body outside the regular body boundaries, locate stimuli to body parts other than those to which the stimuli were applied, to name but a few prominent manifestations. These phenomena can also be associated with certain neurological disorders, notably of the temporo-parietal junction of the cortex. I submit, however, that the relation of experimental and natural body concept aberrations to SMT is ambiguous inasmuch as it is liable to confound basic perceptual-cognitive functions with consciousness.

In passing, one might provocatively say: 'Conscious states can be contagious', referring to the imitation theories of culture [48] and the 'collective states in coupled brains' of which Benzon ( [23], p. 59) speaks as a kind of group intentionality, specifically having the activity the 'musicking' by ensembles in mind. Grasping the intentions of others with one's mirror neuron system seems to be quite generally involved in various forms of interpersonal relations [64, 76, 115].

Returning to basics: the brain's activity responsible for the formation SMT is assumed to consist of complex information processing and representational mechanisms. It is thus indebted to notions adopted from Cognitive Neuroscience [102]. In the next section, I will critically discuss some aspects of the role of Neuro- and Cognitive science for consciousness studies.

3.3 Quality control of adoptions and imports from Neuro- and Cognitive Science

Consciousness studies can be expected to selectively draw on concepts that are based in the Neuro- and Cognitive Sciences. Representational mechanisms and Information processing, mentioned in the context of SMT, are examples of such imports whose standing in the source disciplines is far from being uncontroversial. In pat, the trouble lies there with Informational Semantics [49] and the intuitions it has instilled: it assumes a causal co-variation in the sense that event A carries information about event B, and B has now representational content of A. But law governed processes can virtually be a dime a dozen which makes information ubiquitous, hence the need for subjecting the relation to constraints supplied by an observer. Grush [71, 72] resolves this conundrum by defining representations as "entities which are *used* to stand for something else", with the emphasis on the word 'use', indicating thereby that the proper role of Representations is for off-line use, as he suggests, under counterfactual conditions. Similarly, consider the pattern illustrated by the relation "it is nomically necessary that litmus (s) turns red (r) in acid ": The informational relation obtains between s's being in a certain way, and r's being in a certain way; with the source s being F, and the receptor r being G. Then the fact that r is G carries the information that s is F if and only if it is nomically necessary that s is F given that r is G, subject to *ceteris paribus* conditions which need to be constrained by an observer [118]. Both cases support Haugeland's [74] conclusion that there is no principled way for viewing brain and environment as separate systems, albeit amenable to functional separability by criteria that need to be supplied by an observer and are thus not intrinsic to the system.

Problems with Information and associated notions run still deeper: in Computational and System Neuroscience, data and conclusions are frequently stated in terms of some measure of Information based on Shannon's Mathematical Theory of Communication (MTG), with the offshoots of Information Theory and the notion of Information Processing, the latter largely modeled after the Theory of Computation. The validity of reported data must, thus, depend on their source having met the specific assumptions of the parent Theories: for instance that the neural data fulfill the premises of MTG such as ergodicity and normality of data distribution. However, a recent review and re-appraisal of a large collection of data in the Neuroscience literature shows that they do not generally meet criteria that accord with these assumptions [31, 164]. Instead, there is substantial evidence for fractality and self-similarity in space and time, at all levels of organization, extending from individual neurons to field potentials, EEG and fMRI records, and to perceptual-psychophysical and cognitive functions. Additional recent observational and theoretical studies by Grigolini et al [70] and Allegrini et al 2,3] of global intermittent dynamics of collective excitations in the resting-state Electroencephalogram (EEG) add an important new dimension: the Rapid Transition Processes (RTP), previously identified by Kaplan et al [79] and examined in great detail by Fingelkurts et al [59, 60] are not only evidence for rapid intermittent transition processes of global metastable transitions, but they also display multichannel avalanches in the form of simultaneously occurring RTP's in several EEG recording sites. The avalanches were identified by applying the method which Beggs and Plenz [21] had used at the mesoscopic level of neural organization. Several statistical measures of multichannel avalanches exhibit inverse power-law statistics. Thus, the avalanches attest to a state of self-organized criticality at the level of the whole cortex, and its complexity. On closer analysis, however, it turns out that different cortical areas have different degrees of complexity. The undoubtedly significant consequences of this observation warrant further investigation.

Taken together, these various strands of evidence suggest the need for channeling Theoretical Neuroscience into directions which are at variance with still widely-held beliefs, reflected in currently authoritative sources such as for instance: [40, 42].

Finally, recall that Information does not have a natural Ontology and is accordingly, not intrinsic to the brain: Information, including measures like Entropy etc, refers to the state of knowledge of an observer, and is not intrinsic to the physics of a system [85]. Nor is programmable computation intrinsically a natural process: rather, it is putting Physics in the service of the user's syntax and semantics [166].

Atmanspacher and Rotter [5] direct attention to the intricacies of ontic and epistemic states, the former referring to a physical state as such, as it is independent of an observer; the latter referring to the usually context dependent knowledge that can be obtained about an ontic state. Relations between ontic and epistemic states get intricate in hierarchic inter-level relations : it is then possible that states and properties of a system, viewed epistemically at one level of description can be considered ontic in the perspective of a higher level (i.e. objects constituted from descriptions); also, compositions of lower level objects can be epistemically described, but alternatively also ontically characterized as 'building blocks' of higher level objects [9]. Hence, onticity can be relative to context. Why are these distinctions important ? Because it depends on them whether a given item can be considered on the level of reality and subject to laws of nature, or to subjective perspectival discourse. In complex systems, this distinction is the essential for differentiating (computational) processes intrinsic to ontic states from transactions among their (epistemic) descriptions. These distinctions are also decisive for inter-level relations: for instance, for deriving a description of a system's features from a lower level description, in which case Bishop and Atmanspacher [26] and Atmanspacher [7] additional contingent contextual conditions are required. For illustrating the principle of 'contextual emergence' with an example from Physics, consider that temperature as a novel phenomenal property , by itself alone, cannot be derived from a lower level statistical mechanical description. However, providing a context in the form of certain stability conditions, temperature emerges as a novel property at the level of thermodynamics. In this conceptual framework, coarse grained description of neural activity can provide the necessary conditions for the emergence of (phenomenal) mental state descriptions at the cognitive level, with stability criteria from Symbolic Dynamics as the context [6]. Note that this process establishes an epistemic relation between neuro- and cognitive dynamics, and concurrently also resolves the notorious symbol grounding problem [73].

The point of all this is that confounding observer-relative and intrinsic features of objects or events can become a notorious source of confusion and false attributions, of which certain uses of 'information', 'computing', 'complexity' [10] and 'consciousness' are potentially susceptible victims.

Taken together, these various intrinsic uncertainties in Cognitive Neuroscience and the potential sources of conceptual fallacies point to the merit of the principled generative-constructive approach to consciousness studies, based on brain-based devices in realistic environments, with embodied nervous system models of transparent design whose parameters are under the investigator's control, as examples reviewed in Section 3.1 show. Supplementing judiciously chosen intuitions gleaned from Neurobiology with novel designs and inventions, for stepwise refinement of model performance can be expected to eventually identify optimal design principles and competencies required for approximating the capabilities associated with Consciousness. Take the example of Shanahan's [132] abductive framework for active perception which is another effective way for bypassing the symbol grounding problem [73] at a practical level, though not the principle of the measurement-dependent Heisenberg cut [10] separating the physics of rate-dependent dynamical states from arbitrary symbols [111]; an issue commonly neglected in current Cognitive Neuroscience, yet undermining it at a fundamental level.

With the general outlook of this section in mind, I turn next to proposing new aspects of the 'Relevant Complexity' whose requirement for multidimensionality was the substantive insight of the studies reviewed in Section 3.1

4. Brain State Space Dynamics and Complexity

Studying neural activity in terms of state space representations has proven of immense heuristic value [160]

for capturing the spatial and temporal dynamics of neural systems at micro-, meso- and macroscopic granularity. Its merit lies in making conceptual tools of statistical mechanics [140] available for neural data interpretation. Essentially, it entails associating a measure of neural activity with a point (or region) in a usually higher dimensional space with following the path taken under the control of evolution equations. The evolution can be ontically characterized by Langevin type differential equations, or epistemically as Fokker-Planck type descriptions of point clouds or trajectory bundles [5] In an application of this principle, Allefeld et al [1] examined conditions and criteria for, what they consider, the emergence of mental states from brain electrical dynamics. The State Space approach was also applied by Fell [58] in an attempt to identify neural correlated of consciousness. However, what is of crucial importance for the topic of this essay is the fact that the dynamics of neural systems can display at all levels the property of criticality. By this is meant that the path trajectory can undergo bifurcations due to singularities in certain regions of state space. According to Critical Theory of Statistical Physics, a system's control parameter can be tuned to undergo sudden phase transitions to new macroscopic configurations with distinctly novel properties. One aspect of the system reconfiguration consists of a change of the correlation function among its elements, which characterizes how the value at one point in state space correlates with the value at another point. While ordinarily extending only over short distances, correlation length increases with approach to the critical point where it finally becomes infinite, having established a new pattern of collective interactions among the system's components.

The occurrence of such singularities in neural dynamics and associated state transitions is by now amply established for all levels of neural system organization, and is documented in recent reviews [31, 164]. Brain phase transitions between active-conscious and unconscious states occur at critical values of anesthetic concentration [144, 145]. Their relevance for human cognition is evaluated by Ref. [137]. In the framework of the Global Workspace Model, Dehaene [43] presents evidence for non-conscious local processing within specialized modules prior to reaching a threshold for "global ignition" underlying conscious reportability; the author does not discuss these findings in terms of phase transitions, but his description of observations is highly suggestive of a critical transition in phase space dynamics.

Wallace [161] applied the principle of phase transition to the Global Workspace Model (GWM) of Baars (see Section 3.1), viewing consciousness associated with a core of limited processing capacity which receives input from a collection of distributed, specialized processors and (unconscious) contexts. Selecting classical and semiclassical results from network theory [108], Wallace considers network nodes as processors of the global workspace model, and network links as mutual information between them . He then for explores conditions under which the network coalesces by phase transition to a 'giant component': a term taken from Percolation Theory [142]; see also an illustration in [166] to signify the merger of network nodes to one large assembly. Intuitively, one may interpret the formation of the Giant complex as a cognitive event. The model illustrates that network topology and clustering of linkages are tunable parameters of network configurations that are interpretable as descriptors of psychological concepts. This model offers the advantage of being intuitively better accessible, and bearing more directly on data from biological network- and fMRI studies than does an earlier version which will be referred to in Section 5.

Although not being part of Wallace's own discussion of the model, it does invite one to speculate on underlying neural mechanisms of the kind studied in Ref. [88] as phase transitions in neuro-percolation models. Percolation also turned up in the theoretical work of Ref. [29] on entirely different grounds and in the context of different premises: applying field theoretic methods to non-equilibrium statistical processes in Markovian neural networks, the authors showed the proclivity of their computational models for dynamical phase transitions of the (directed) percolation type.

This dynamics of state phase transitions is considered universal in its independence of details at the microscopic level [140, 136] metaphorically speaking, the system 'forgets' its original microscopic structure and properties by coarse-graining, and loses all characteristic length scales for system specific variables: it becomes scale invariant, i.e. fractal [141]. Dissipative complex systems can attain the state of criticality by self-organization where the critical state is an attractor for the dynamics [15]. Applying this theoretical framework, Kitzbichler et al [83] compared phase synchronization and scaling in critical models with global synchronization in fMRI and MEG records from humans at rest, and found essential correspondence, suggesting that functional systems in brain exist in a state of endogenous criticality. Fraiman et al [62] determined brain correlation networks from fMRI voxel-to-voxel correlations under the same computational conditions. Contrasting their human data with correlation networks from computational simulation of a computational model of magnetic spins (Ising model) showed close correspondence in all relevant respects, thus also supporting the conjecture of the human brain at rest functioning near a critical point. In addition, the investigators of this and Ref. [56] were struck by the emergence of nontrivial collective states in the critical state. As noted before, state transition in physical systems is reflected by the correlation coefficient at two different points in space and at different times. In their analyses of human data at successive spatial coarse-graining steps, they uncovered self-similarity of correlations, while the temporal pattern followed as usual 1/f frequency (i.e. power-law) behavior of the power spectrum. The authors of this study suggest that the dynamical self-similarity affords the opportunity for brain states to alternate between metastable states (attractors) of predominantly short- or long distance correlations. Phenomenologically, this suggestion is reminiscent of the brain's state according to the theory of Extended Coordination Dynamics which, however, interprets Metastability in a differently: namely as expression of the opposing tendencies of complementary pairs of brain states, yet functioning collectively as coupled pairs, without ever becoming attractors for taking control [55, 81, 82]. Although this is not so stated by the authors, one could perhaps attribute the 'overlapping tendencies' to the sharing of an attractor basin ? In either case, Metastability appears to subserve local segregation (specialization) and global integration of cortical regions.

The virtual ubiquity of fractal phenomenology in the brain and of many cognitive functions [24, 70, 164], in part at least accounted for by self-organizing processes, is a telling signature of their complexity, and supports viewing the brain poised to criticality where inverse power-law correlations obtain in space and time. To this, Allegrini et al [2] added important new evidence, based on studying the 'rapid transition processes' (RTP) in the Electroencephalogram, originally identified by Kaplan et al [79] these are abrupt changes in EEG amplltude, interspersed between segments of regular amplitude and wave shape. When recording from several EEG electrodes concurrently, Figelkurts et al [59, 60] had noted simultaneity of RTP's in several recording channels at a frequency above the expected statistical average, suggesting collective activity of functional neuron assemblies. To isolate concurrent spatio-temporal patterns of RTP's. Allegrini (l.c.) then applied the procedure of Ref. [21] had successfully used for avalanche detection at the mesoscopic level. The results unequivocally identified a fractal avalanching process of self-organized criticality involved in global metastable transitions. Moreover, regional differences in scaling behavior showed that cortical areas differ in respect to their complexity. We are left with the question: do these data characterize the status of the default -mode network (the subjects were at rest), or do they speak to the integrated neural dynamics sustaining consciousness?

As is well known, Complexity (together with emergence, information dynamics, and related notions) is a bottomless barrel of different opinions, idiosyncratic definitions, and perceptions: see for instance: Ref. [113]. However, the focus of interest has shifted towards the nascent field of Complex Networks, situated at the intersection of graph theory and statistical physics [36]. Their common features are inverse power-law statistical distributions, multiplicity of scales, manifestations of non-stationary, and non-ergodic statistical processes. A complex network's capacity for storing information is defined by its statistical entropy [4, 134]. Real networks occurring in Nature seem to cluster in a relatively narrow region of entropy-noise space . Internal to a network, information transfer between network nodes depends on the network topology: it is optimal at or near the network's  phase transition  [80]. At this critical point, self-organized neural networks are also optimized for cooperative learning [41]  and memory function [94; see also [28], section 5.3].

Although also dealing with information transfer, Allegrini's (II.c.) finding of  cortical areas differing in complexity (see foregoing) introduces an  entirely different situation, namely information exchange <u>between</u> complex networks. A series of theoretical-mathematical studies, summarized by West et al [171] were concerned with analyzing how one complex network responds to perturbation by a second complex network: this is viewed as a form of information transport from one network to another, in principle comparable to stochastic resonance (for instance: Moss et al[106], except for  involving interacting complex networks. In this case, the interaction of complex System S with complex system P shows that P inherits the correlation function from S;  it does so optimally if S and P have the same power-law index; hence the designation of this phenomenon as Complexity Matching Effect. Allegrini et al (l.c.) and Bianconi et al (l.c.) validated  the theoretical predictions with data from Electroencephalogram recordings, statistically  representing a non-ergodic, non-Poisson renewal process, with power-law indices <2  complexity. Clearly, this approach introduces an entirely new perspective for interpreting and evaluating neurophysiological data at the complex systems level, and challenges prevailing notions of 'information flow'  in nervous systems. Applying the theory of the Complexity Matching Effect to dynamic interactions between central core and modules of the Global Workspace Theory could yield some revealing insights.

The multiplicity of scales encountered in complex networks invites the application of Renormalization Group Theory, a set of concepts and methods allowing one to understand phenomena in many fields of physics including classical statistical mechanics of non-equilibrium systems. It proves particularly useful to understand phenomena where fluctuations involving many scales of length and time scales lead to the emergence of new collective behavior in complex systems. It is the topic of the next Section.

5. Applying the Renormalization Group Theory (RNG)

RNG is both a computational approach and a way of viewing reality. In the former, the Renormalization  Group Theory studies the transformational dynamics by which a complex system space maps onto itself in the stepwise traversal of the basin of attraction (the critical manifold)  towards a critical fixed point.  On its path, the system 'defines many worlds' (as Kadanoff  [77] aptly put it). The consecutive 'worlds' originate by phase transitions which occur stepwise along the trajectory of the RNG  transformation.  They are at the microscopic level its stepwise generated self-similar solutions, each reflecting intermediate asymptotic behavior [17]  and  increasing range of correlations among constituents. Viewing them as 'different worlds' is due to the fact that the  new micro-level correlation  patterns express themselves at  the macroscopic level as qualitatively different novelties, with their own new laws and new descriptors:  think, for instance, of the phase transition from water to ice.  At each step, the essential aspect of the transformation consists in progressive coarse-graining and change in length scale: coarse-graining being the coupling  microscopic degrees of freedom  to each other to make them  effectively act as  single entities with correlation lengths on all lengths, implying scale invariance. In effect, RNG eliminates microscopic details that are inessential or irrelevant for determining the system's behavior at the critical point ( [20],  Ch. 4).  This accounts for the Multiple Realizability  of  complex systems: different  microscopic constituents and dynamics giving rise to identical macroscopic behavior. At the critical point, the system's correlation coefficients become infinite, reflecting the enormous number of correlations among constituents.  At this point, the 'critical exponents' are universal in the sense that they apply to entire classes of systems which can consist of materials of very different microscopic constituents. One then speaks of universality classes that have some very general system properties in common, such as dimensionality and the length of range of interactions. (For introductory accounts: [68, 173]).

For the purposes of this essay, the focus is on the "Renormalization Group View of the World" [133]. It derives from the computational approach of RNG inasmuch as it elevates its computational achievements to an epistemic principle: the punctuated phase transitions along consecutive steps on the trajectory of RNG delimit levels of qualitatively different realities, each on its own scale and grain of resolution, and governed by its own laws. Although unrelated to RNG, consider for intuitive appreciation of this phenomenon the following historically notable illustration: when Anthony von Leuwenhoek in 1674 increased the magnification of his microscope to a certain (critical !) degree, he saw suddenly a populated world of protists, never previously expected to be part of the water drop he examined. In other words, he accessed a new level of reality which exhibited its own ontology and the laws that govern it. In RNG terms, his changing the magnification is equivalent to a step along a transformation trajectory associated with the change of the grain of observation; in this case to a finer grain. This much for a stunning entry into a new, previously unknown level of Reality. More commonly in Physics, one is interested in the opposite direction, which is the case of collective emergence, associated with changing to a coarser grain with elements at the micro-level coalescing to larger entities (as if becoming myopic). But the basic issue is this: like a deck of cards, reality is considered a hierarchy of levels where, on each level, elements are collectively organized to structures with their own laws, properties, intrinsic scales, and arrangement (correlations) among its elements; in short: each level is an ontology, as such requiring its own, distinct description. Conversely, two levels of description will have radically different ontologies. Two consecutive levels are related to one another by sharp phase transitions. In effect, the transition from a lower to a higher level organizes the distinct objects of macroscopic levels [92]. The qualitative novelties of consecutive levels arise *de novo* in virtue of the system dynamics, and do not stand in any logical relation to one another, hence are not logically deducible from one another. For instance, our (non-fundamental) ontology of everyday objects is in this framework ultimately an intermediate asymptotic approximation (see above) to the more fundamental ontology of quarks and electrons. An analogous relation obtains between Theories in Physics [78]: as example, the ontology of Newtonian mechanics may be viewed as a coarse grained version of the more fundamental general-relativistic ontology ([133], p. 40).

In an application of this conceptual framework to cognition and by implication to consciousness, Wallace [162] outlined the sketch of a theoretical scheme that utilizes some central ideas of information theory in the context of a renormalization point of view. Reminiscent of the spirit of the Global Neuronal Workspace, cognitive modules, here constituted basically as language like structures, would interact by punctuated phase transitions to generate large coherent structures, locking under renormalization at a critical point.

The objective of this essay, outlined in the Introduction, is to conceive the realm of the Subjective as an ontology whose epistemic access constitutes the features of experiential subjectivity. In the preceding sections, several benchmarks are identified which such an approach will have to pass: criticality of Neurodynamics of a nervous system built by environmental specification (in the spirit of the Darwinian model), accommodation to aspects of body mechanics and dynamics in the interaction with a bounded region of the environment, and an adequate account of the body's internal state and homeostatic regulations. I suggest that these different ingredients, taken together, constitute (at least in part) the multidimensionality of what Seth et al [128] seem to have in mind as 'relevant complexity'.

Circumnavigating in the foregoing Sections what appear some conceptual hurdles and flawed conclusions from experimental data and theoretical studies, I arrive at an alternative to prior attempts of rationalizing subjectivity in Natural Science terms: namely, that the Renormalization world view is useful and plausible conceptual platform. Specifically, I envision an ontology of Personhood which, in the format of a complex network, encompasses a representation of the body (somatic as well as autonomic), in association with the brain, and portions of the (cognitive) environment. The domain of the Subjective is then thought to arise as a complex network along a trajectory of phase transitions along a path to the fixed point which marks the fully conscious state. The sequence of consecutive phase transitions would constitute distinct levels whereby each of the consecutive levels of the Renormalization process presents a collective achievement with different granularity of resolution in details and different degrees of correlation. The density of correlation among the elements of a level reaches a maximum at the Fixed Point of RNG where it covers the total range of self-similar (fractal) scales, indicative of optimal integration of correlated activity. Note the difference to the more common and more intuitive view that Consciousness 'emerges' from neural mechanisms as a higher level of organization. The opposite is proposed here: The levels of subjectivity arise as ontologies in the phase transitions from the more encompassing (higher level) ontology of the world-body-brain Physics so subordinate levels; and can do so stepwise at consecutive levels of different resolution in details. This may be associated with the taxonomy of conscious, preconscious and subliminal processing which Dehaene ( l.c. ) describes. As an added bonus, this approach eliminates the notorious problem of multiple realizability ([20], Ch.5.2 ) since it is an integral aspect of RNG that many different configurations at one level can become identical at another level, on account of coarse-graining.

Although drawn in crude brushstrokes, this sketch of applying the Renormalization view of the world to Consciousness is in principle amenable to testing its plausibility. This can be pursued by capitalizing in models on the computational assets of RNG. Refs. [135], [119] and [114] are suitable starting points for studying RNG transformations of complex networks and offer valuable clues.

As a final implication of Renormalization Group application, note that the forgoing speculations situate the problem of consciousness in the parent domain of Condense Matter Physics (see for instance: [95]). This suggests an interesting perspective: for, as stated in the introductory paragraph of his section, the principle of Universality would permit identifying whether, and to what degree, other materials in Nature may possess features indicative of a form of consciousness, once it would have been possible to determine the Universality class of which the ontology of Personhood is a member.

Summary.

The review of the principal current approaches to modeling cognition and consciousness leads me to suggest that none of them possesses the degree of multidimensionality or 'Relevant Complexity' (l.c.) , required for emulating more than, at best, one or another feature of primary consciousness to an elementary degree. Moreover, I identify and discuss what I perceive to be basic conceptual flaws in some conventional assumptions that Cognitive Neuroscience and Neurophysiology bring to the task. Nor are the basic insights of phase space dynamics and criticality adequately taken into account. In the light of this situation, I propose a different approach: first, consider the realm of Subjectivity (as hallmark of consciousness) as an ontology in its own right (as Searle suggested in 1992); and, second, think in terms of the Renormalization Group's view of Reality: the subjectivity of Consciousness is then the epistemic interpretation of the ontology that originates in phase transitions of the brain-body-environment system.